\documentclass{article}

\usepackage{amsthm}  

\newtheorem{theorem}{\bf Theorem}[section]

\newtheorem*{definition}{\bf Definition}
\newtheorem{remark}{\it Remark}[section]

\usepackage{float}

\usepackage{xcolor}
\usepackage{mdframed}
\newmdenv[
  backgroundcolor=gray!4,
  linecolor=gray!60,
  roundcorner=15pt,
  linewidth=1pt,
  frametitlebackgroundcolor=gray!15,
  frametitlefont=\bfseries\color{black},
  skipabove=10pt,
  skipbelow=10pt
]{mybox}

\usepackage{bm}
\usepackage{comment}
\usepackage{subcaption}

\usepackage{arxiv}
\usepackage{xcolor}
\usepackage{mdframed}

\definecolor{dgray}{rgb}{0.4, 0.4, 0.4}
\newcommand\dgray{\color{dgray}}

\newcommand{\variance}{\sigma^2}

\newcommand{\thetav}{\theta}

\newcommand{\Prob}{\mathbb{P}}

\newcommand{\D}{\mathcal D}
\newcommand{\Bres}{\text{B-res}}
\newcommand{\ppd}{\text{PPD}}
\usepackage{natbib}
\usepackage{bm}
\usepackage{comment}
\usepackage{subcaption}
\usepackage[utf8]{inputenc} 
\usepackage[T1]{fontenc}    
\usepackage{hyperref}       
\usepackage{url}            
\usepackage{booktabs}       
\usepackage{amsfonts}       
\usepackage{nicefrac}       
\usepackage{microtype}      
\usepackage{lipsum}
\usepackage{graphicx}
\graphicspath{ {./images/} }
\usepackage{amsmath}
\usepackage{amssymb}
\usepackage{amsthm}

\title{The Interplay between Bayesian Inference and Conformal Prediction
}

\author{ \href{https://orcid.org/0000-0003-2501-8795}{\includegraphics[scale=0.06]{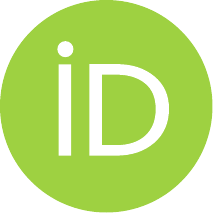}\hspace{1mm}Nina Deliu}\\
	MEMOTEF, Sapienza Università di Roma (IT)\\
        MRC – Biostatistics Unit, University of Cambridge (UK) \\
        \texttt{nina.deliu@uniroma1.it} \\
	\And
	\href{https://orcid.org/0000-0003-2089-3371}{\includegraphics[scale=0.06]{orcid.pdf}\hspace{1mm}Brunero Liseo} \\
	MEMOTEF, Sapienza Università di Roma (IT)\\
	\texttt{brunero.liseo@uniroma1.it}\\
}

\date{}

\begin{document}
\maketitle
\begin{abstract}
Conformal prediction has emerged as a cutting-edge methodology in statistics and machine learning, providing prediction intervals with finite-sample frequentist coverage guarantees. Yet, its interplay with Bayesian statistics--often criticised for lacking frequentist guarantees--remains underexplored. Recent work has suggested that conformal prediction can serve to ``calibrate'' Bayesian credible sets, thereby imparting frequentist validity and motivating deeper investigation into frequentist–Bayesian hybrids. We further argue that Bayesian procedures have the potential to enhance conformal prediction, not only in terms of more informative intervals, but also for achieving nearly optimal solutions under a decision-theoretic framework. Thus, the two paradigms can be jointly used for a principled balance between validity and efficiency. This work provides a basis for bridging this gap. After surveying existing ideas, we formalise the Bayesian conformal inference framework, covering challenging aspects such as statistical efficiency and computational complexity.

\end{abstract}


\section{Introduction} \label{sec: intro}

The history of statistics has long been marked by debates between Bayesian and frequentist schools of thought; see \cite{piccinato1992critical} and \cite{efron2005bayesians}. Each tradition carries its own philosophical foundations: the Bayesian paradigm is grounded in probabilistic updating of beliefs through prior-to-posterior inference and obeys the likelihood principle, while the frequentist paradigm relies on procedures that maintain long-run error guarantees across repeated sampling under fixed conditions (a.k.a. ``procedural frequentist principle'', as outlined by Berger~\cite{berger_2023}). The foundational tension is unlikely to be resolved, nor, perhaps, should it be; yet, both at a methodological and applied level, the boundaries are more and more permeated and shaped by practical considerations.

\maketitle

Among different, very radical approaches, a more conciliatory position has always remained alive. This sought to guarantee, on the one hand, the conditional properties of the procedures and, on the other, the long-run properties. Important works supporting this position can be found in \cite{rubin1984} and \cite{bayarriberger2004}, or in \cite{berger1999, berger2003could} for the problem of testing hypotheses. Scholars are increasingly arguing for an interplay rather than for the superiority of one paradigm: as Bayarri and Berger (2004, p. 58) observed, \emph{``each approach has a great deal to contribute to statistical practice and each is actually essential for full development of the other approach''}~\cite{bayarriberger2004}. This research stream has given rise to hybrid approaches such as empirical Bayes~\cite{carlin2000bayes}, calibrated Bayes~\cite{little2006calibrated, little2011}, matching priors~\cite{datta2004probability}, 
or priors that guarantee \textit{admissibility}~\cite{means1993choice}, among others.

The modern landscape of statistics and machine learning, where {\it prediction} is acquiring increasing attention, pushes further the need for a practical reconciliation of the two perspectives. Uncertainty quantification occupies a central position in this reconciliation. Undoubtedly, Bayesian statistics remains a cornerstone in representing and propagating uncertainty from parameters to future observations: it does so via entire probability structures in the form of prior/posterior {\it predictive} distributions. These are commonly summarised through Bayesian intervals in the form of highest-posterior predictive density (HPPD) intervals~\cite{box_bayesian_1992,deliu2024alternative}. Although capable of incorporating prior knowledge and adapting flexibly to complex structures, posterior predictive intervals may fail to possess coverage guarantees under a long-run frequentist perspective. A prediction procedure is said to ensure frequentist coverage guarantee if the actual long-run coverage of the predictive region matches (or exceeds) the nominal coverage $1-\alpha$, with $\alpha \in (0,1)$. Formally, let $Y_1, \dots, Y_n \sim P$ denote the sample data, with $P \in \mathcal{P}$ an unknown distribution, and let $Y_{n+1} \sim P$ be a future observation; for simplicity, assume $Y_i \in \mathbb{R},\ i = 1,\dots, n+1$. Let $\mathcal{C}_{n,1-\alpha} = \mathcal{C}_{1-\alpha}(Y_1, \dots, Y_n)$ denote a {\it prediction interval} constructed at the nominal level $1-\alpha$, on the basis of the sample data. The procedure $\mathcal{C}_{n,1-\alpha}$ is said to satisfy frequentist coverage guarantees (or {\it validity}) if
\begin{align}\label{eq: freq_cov}
    \Prob_P (\mathcal{C}_{n,1-\alpha}) \doteq \Prob_{Y_1,\dots,Y_{n+1} \sim P}
    \Big(Y_{n+1} \in \mathcal{C}_{n,1-\alpha} \Big) \geq 1-\alpha,
    \quad \forall P \in \mathcal{P}.
\end{align}
Note that the probability in Eq.~\eqref{eq: freq_cov} is taken jointly over the randomness of the sample data and the new observation, reflecting the hypothetical repeated sampling or long-run philosophy. Whereas, for a $1-\alpha$ Bayesian HPPD interval, say $\mathcal{C}^{\text{HPPD}}_{n,1-\alpha}$, the probability is conditional on the already observed data and holds under the posterior predictive distribution $p(\cdot | Y_1,\dots,Y_{n})$, i.e., 
\begin{align}\label{eq: Bay_cov}
    \Prob_{Y_{n+1} \sim p(\cdot | Y_1,\dots,Y_{n})} \Big(Y_{n+1} \in \mathcal{C}^{\text{HPPD}}_{n,1-\alpha} | {Y_1,\dots,Y_{n}}\Big) \geq 1-\alpha.
\end{align}

The coverage principle \eqref{eq: freq_cov} has clear implications for Bayesian practice: intervals that fail to attain nominal coverage risk to become unreliable. Critics point out that Bayesian intervals may easily undercover or overcover, unless the model is correctly specified, which is rarely ensured in practice. This gap can be substantial in nonlinear, high-dimensional, or nonparametric problems and is exacerbated when using strong prior information. Therefore, assigning probabilistic interpretations to intervals lacking frequentist validity is not merely a technical oversight, but a form of misrepresentation that bears serious consequences for scientific integrity and public trust~\cite{fraser2011, wasserman2011}. The stakes are highest in regulatory applications, such as FDA-monitored clinical trials, where procedures must deliver explicit (frequentist) error control, even under Bayesian designs~\cite{fda_adaptive_2019}. Nonetheless, in government and public policy, even where a certain form of ``objectivity'' is deemed necessary, the benefits of a Bayesian approach have, over time, been well introduced into statistical practice~\cite{fienberg2011bayesian}. 

Conformal prediction (CP)~\cite{vovk2005algorithmic,lei2018distribution} has recently emerged as a promising tool in this direction. Under the sole assumption of exchangeability among observations, conformal methods can yield finite-sample frequentist coverage guarantees for essentially any prediction method -- be it parametric, nonparametric, Bayesian, or not. This is true regardless of the working model and the prior. Early work prompted the investigation of using conformal prediction (via a test-inversion procedure) in order to ``frequentise'' Bayesian intervals~\cite{wasserman2011}. More recent contributions have further expanded this theme, exploring the so-called {\it full conformal prediction} to ``conformalise'' Bayes~\cite{fong2021} or to achieve a principled ``frequentist and Bayesian'' (FAB) compromise~\cite{hoff2023,bersson2024}. 
Notably, Hoff and colleagues \cite{hoff2023,bersson2024} point to the potential of embedding indirect or prior information to construct a prediction region that not only maintains a target frequentist coverage, but also achieves forms of optimality. The context of small areas considered by \cite{bersson2024} is paradigmatic in this sense. Small-area estimation is the most obvious example of how much a Bayesian approach is necessary even in ``official'' contexts, where the problem of the frequently small sample size of individual areas is overcome by the hierarchical approach that allows ``borrowing strength'' \citep{datta1991}. If the introduction of extra-experimental information in such contexts can endanger the validity, a CP-based approach avoids this eventuality.

These observations point towards a deeper level of synthesis. If, on the one hand, CP provides the scaffolding for validity, on the other hand, Bayesian principles can flexibly leverage probabilistic structures to sharpen the expected width or volume of the region, shaping {\it efficiency}. Often, conformal methods may produce intervals that are conservative or insufficiently adaptive to the data structure, creating an opening for Bayesian methods. In this light, conformal prediction and Bayesian inference can complement each other to mutually reinforce their properties. 

The gist of this paper is to contribute to the growing literature on CP, highlighting its interplay with Bayesian inference. A major emphasis is placed on the operational roles of efficiency and computational complexity, made especially salient by the dual nature of {\it full} CP and Bayesian approaches. After introducing CP in Section~\ref{sec: CP}, in Section~\ref{sec: CP_Bayes}, we offer a unified overview of recent Bayesian developments: we discuss scalable alternatives within {\it full} CP and formalise a Bayesian variant of {\it split} CP, proposing and evaluating different Bayesian conformity measures. 

Our aim is certainly not to resolve the long-standing Bayesian-frequentist debate. More humbly, we believe the use of conformal methods can contribute to adjusting the prediction 
provided by those Bayesian methodologies that have been exposed
as having potentially serious frequentist problems \cite{bayarriberger2004}. In a spirit of modern statistics, based less on division and more on synthesis, Bayesian conformal inference may be viewed as one concrete pathway.

\section{Conformal Prediction} \label{sec: CP}

Conformal prediction represents a flexible yet robust framework that is attracting considerable attention in modern statistics. It is, in principle, used for quantifying the uncertainty in predictions made by any arbitrary strategy, and its underlying idea is quite simple: the basic theory stems from the relationship among exchangeable random variables, rank statistics, and sample quantiles. Specifically, if $R_1,\dots, R_n$ and $R_{n+1}$ are exchangeable (e.g., i.i.d.) realisations of a scalar random variable, then their rank is uniformly distributed over $\{1, \dots, n+1\}$. Furthermore, denoted by $q_{n, 1-\alpha} \doteq R_{(\lceil (n+1)(1-\alpha) \rceil)}$ the $1-\alpha$ sample quantile of $R_1,\dots, R_{n+1}$, with $R_{(j)}$ being the $j$-th ordered sample element and $\lceil \cdot \rceil$ the ceiling function, the following holds for any finite sample size $n$ and $\alpha \in (0,1)$:
\begin{align} \label{eq: sample_q}
    \Prob \left(R_{n+1} \leq q_{n, 1-\alpha}\right) \geq 1-\alpha.
\end{align}

\paragraph{Setup and notation.} Let $\D_n = \{Z_i = (\bm{X}_i, Y_i)\}_{i = 1}^n$ be a labelled sample of size $n$, with $\bm{X}_i \in \mathcal{X} \subset \mathbb{R}^p$, for an integer $p \geq 1$, and $Y_i \in \mathcal{Y} \subset \mathbb{R}$ denoting the covariate set and the response variable of interest, respectively. Consider a new observation indexed by $n+1$ with feature $\bm{X}_{n+1}$; the interest is in quantifying the uncertainty in the associated (unobserved) response $Y_{n+1}$ via a $(1-\alpha)$-level prediction interval $\mathcal{C}_{n, 1-\alpha}(\bm{X}_{n+1})$. We want $\mathcal{C}_{n, 1-\alpha}(\bm{X}_{n+1})$ to satisfy the coverage guarantee in Eq.~\eqref{eq: freq_cov}. By assuming exchangeability among the pairs $Z_i = (\bm{X}_i, Y_i),\ i = 1,\dots, n+1$, this property can be easily achieved under a CP framework, fundamentally based on Eq.~\eqref{eq: sample_q}. 

One of the main features of CP, compared to e.g., asymptotic prediction intervals, resampling methods, or a mere adoption of the quantile property in Eq.~\eqref{eq: sample_q}, is the use of a \textit{(non-)conformity score function} $r\!: \mathcal{X} \times \mathcal{Y} \to \mathbb{R}$ to quantify the (dis-)similarity or {\it (non-)conformity} of any point $Z = (\bm{X}, Y)$ to an observed sample $\D$. An important property of any conformity function is that it is symmetric in $\D$, i.e., the conformity function for $Z$ is invariant to any permutation of the elements of $\D$. By leveraging the symmetry of $r$ together with the exchangeability of the $Z_i$'s, conformity scores
\begin{align*}
    R_i = r(Z_i; \D),\quad i = 1,2\dots,
\end{align*}
are themselves exchangeable, allowing Eq.~\eqref{eq: sample_q} to be directly applied to build a valid $\mathcal{C}_{n, 1-\alpha}$. We note that the notation ``;'' in $r(z; \D)$ does not reflect a conditioning statement: it quantifies the (dis-)agreement of $z$ to the patterns observed in the sample $\D$, which plays the role of a ``training'' set. Different CP frameworks have been developed according to the ways exchangeability among scores is ensured; this often involves an accurate choice of the training set $\D$. We now outline the two main ones: \emph{full conformal prediction} ({\it full} CP) and \emph{split conformal prediction} ({\it split} CP).

\paragraph{Full CP.} The original framework is rooted in what is today referred to as {\it full conformal prediction}~\cite{vovk2005algorithmic}. Although less practical due to its computational complexity, it nonetheless remains an elegant and robust methodology, often leading to more efficient results. 
The idea is to use the full sample both for ``training'' a prediction model and for calculating the critical sample quantile in Eq.~\eqref{eq: sample_q}. Denote by $Z^y_{n+1} \doteq (\bm{X}_{n+1},y)$ a new future point, where $y$ is a suitable {\it candidate} for $Y_{n+1}$, and let $\D^y_{n+1} \doteq \D_n \cup \{Z^y_{n+1}\}$ be an {\it augmented} sample set. The $100(1-\alpha)\%$ full CP interval, say $\mathcal{C}^{\text{full}}_{n, 1-\alpha}$, for the response of the new unit can be obtained as:
\begin{align} \label{eq: CP_full}
   \mathcal{C}^{\text{full}}_{n, 1-\alpha}(\bm X_{n+1}) = \{y\!: R^y_{n+1} \leq R_{(\lceil (n+1)(1-\alpha) \rceil)} \ \text{among}\ \{R_i\}_{i = 1}^n\},
\end{align}
where $R_{i} = r(Z_{i}; \D^y_{n+1})$, for $i = 1,\dots, n$, and $R^y_{n+1} = r(Z^y_{n+1}; \D^y_{n+1})$, are {\it non-conformity} scores. In fact, the smaller $R^y_{n+1}$, the more likely it is for $y$ to belong to the set. The so-called {\it augmented} set $\D^y_{n+1}$ ensures symmetry of $r$ across all the observed $Y_i$'s as well as the non-observed candidate $y$, preserving exchangeability across all scores $R_i\; i = 1,\dots,n$, and $R^y_{n+1}$. In essence, the construction of a full CP interval involves evaluating the inclusion or acceptance of a set of candidates $y \in \mathcal{Y}^{\text{grid}} \subseteq \mathbb{R}$, where $\mathcal{Y}^{\text{grid}}$ is a reasonable grid for the variable of interest $Y$. The inclusion of $y$ is determined according to the rank of its conformity score $R^y_{n+1}$ compared to those of the observed sample scores; for a {\it non-conformity} score, this means
 \begin{align*}
    y \in \mathcal{C}^{\text{full}}_{n, 1-\alpha}(\bm X_{n+1})\quad  \iff \quad R^y_{n+1} \leq q_{n, 1-\alpha} = R_{(\lceil (n+1)(1-\alpha) \rceil)}\ \text{among}\ \{R_i\}_{i =1}^n.
\end{align*}

\paragraph{Split CP.} This represents the most popular and practical framework in CP, due to its substantial advantages in terms of memory and computation expense. The concrete idea can be summarised as follows. First, the sample $\D_n$ is partitioned into two subsets of approximately the same sizes $n_T$ and $n_C$, with $n_T+n_C = n$, called \textit{train set} and \textit{calibration set}, respectively; that is,
\begin{align*}
    \D_n = \D_{n_T}^{\texttt{Train}}  \cup \D_{n_C}^{\texttt{Cal}},\quad \quad \text{with}\ \ \ \D_{n_T}^{\texttt{Train}}  \cap \D_{n_C}^{\texttt{Cal}} = \emptyset.
\end{align*}
The first set $\D_{n_T}^{\texttt{Train}}$ is used to fit or train a {\it point} predictor $\hat{f}_{n_T}\!: \mathcal{X} \to \mathcal{Y}$, while the second set $\D_{n_C}^{\texttt{Cal}}$ has the key role of computing a set of $n_C$ calibration scores $R_i = r(Z_i; \D_{n_T}^{\texttt{Train}}),\ i \in \D_{n_C}^{\texttt{Cal}}$, and a valid quantile $q_{n_C, 1-\alpha}$. The validity of the quantile is ensured by the split procedure: having observed only the sample points in $\D_{n_T}^{\texttt{Train}}$, all the points in the calibration set, as well as the future unit $Z_{n+1} = (\bm{X}_{n+1}, Y_{n+1})$, are treated equally, as none of these is used in the fitting procedure. In this way, exchangeability is preserved. Using a {\it non-conformity} score function $r$, the $100(1-\alpha)\%$ CP interval, say $\mathcal{C}^{\text{split}}_{n, 1-\alpha}$, for the response of a new unit with covariates $\bm{X}_{n+1}$ is obtained as:
\begin{align} \label{eq: CP_split}
   \mathcal{C}^{\text{split}}_{n, 1-\alpha}(\bm X_{n+1}) = \left[\hat{f}_{n_T}(\bm X_{n+1}) - q_{n_C, 1-\alpha},\  \hat{f}_{n_T}(\bm X_{n+1}) + q_{n_C, 1-\alpha}\right],
\end{align}
where $q_{n_C, 1-\alpha} = R_{(\lceil (n_C+1)(1-\alpha) \rceil)}$ amongst the set $\{R_i; \, i \in \D_{n_C}^{\texttt{Cal}}\}$. The approach closely resembles a validation-set or holdout approach, but it further enjoys the finite-sample guarantee in Eq.~\eqref{eq: sample_q} due to the sample quantile adjustment for a future unit (herein, {\it conformal quantile}). 

\section{Toward a Bayesian Conformal Prediction Framework} \label{sec: CP_Bayes}

The conformity function $r$ is a central component of CP, acting as the quantitative basis for assessing candidate predictions $y$ against observed data $\D$. In a regression (continuous) setting, a common choice is the absolute fitted residual, that is, $r(z; \D) = \big|y - \hat{f}(\bm{x})\big|$, where $z = (\bm{x}, y)$ and $\hat{f}$ is a point estimator of the underlying regression model fitted on $\D$. This yields prediction intervals centred around the fitted regression function, with constant width. Variants of this measure have been developed to enhance local adaptivity, for instance by locally scaling residuals to account for heteroscedasticity or covariate-dependent variability~\cite{romano2019conformalized, lei2018distribution}. In the classification (discrete) case, conformity functions are typically based on the probability assigned to each of the classes in the form of likelihood or cumulative likelihood scores~\cite{romano2020classification}. For example, denoted by $\widehat{p}(y \mid \bm{x})$ a probabilistic classifier that estimates $\mathbb{P}(Y = y \mid \bm{X} = \bm{x})$, for $y = 1,\dots,K$, with $K$ the number of classes, one may directly define $r(z; \D) = \widehat{p}(y \mid \bm{x})$, so that more probable classes are deemed more conforming. 

From a Bayesian perspective, the most natural bridge to CP lies in the choice of a Bayesian conformity score function. Since the goal is predictive inference, a principled candidate is the {\it posterior predictive distribution} (PPD), which directly reflects the plausibility of an outcome $y$ for $Y_{n+1}$ at a new $\bm{X}_{n+1} = \bm{x}_{n+1}$, in light of the observed data $\D$ and the postulated likelihood model and prior. Given, e.g., a parametric model $f(y \mid \bm{x}, \theta)$, with $\theta \in \Theta$ a (scalar or vector) parameter, and a prior $\pi(\theta)$ on the parameter, the PPD is given by:
\begin{align} \label{eq: ppd}
    p(y \mid \bm{x}_{n+1}, \D) =  \int_{\Theta} f(y \mid \bm{x}_{n+1}, \theta)\, \pi(\theta \mid \D)\, d\theta \propto \int_{\Theta} f(y \mid \bm{x}_{n+1}, \theta)\, \prod_{i=1}^{|\D|} f(y_i \mid \bm{x}_i, \theta)\, \pi(\theta)\, d\theta,
\end{align}
where $\pi(\theta \mid \D)$ represents the posterior distribution for $\theta$, and $|\D|$ is the cardinality of the set $\D$.

Using the PPD as a conformity score allows Bayesian HPPD intervals to inherit finite-sample validity: CP-based approaches provide the frequentist coverage guarantee, while the Bayesian component contributes probabilistic structure that can adapt to model complexity, incorporate prior information, and exploit hierarchical representations. In this sense, conformalisation and Bayesian modelling play complementary roles: the conformal step ensures coverage regardless of model misspecification, and the Bayesian layer enhances efficiency, sharpness, and flexibility in capturing data-generating mechanisms~\citep{wasserman2011,vovk2005algorithmic}. 

In what follows, we outline the procedural Bayesian CP framework, highlighting the principal challenges that arise throughout implementation and discussing approaches for resolution.

\subsection{Bayesian Full CP} \label{sec: Bayes_fullCP}

As illustrated in Section~\ref{sec: CP}, full CP uses the entire observed sample (augmented with a candidate point $y \in \mathcal{Y}^{\text{grid}}$ to be assessed for inclusion in the prediction interval) in the training step. Given an exchangeable sample $\D_n$ of $n$ observations and a candidate $y$ for a future observation $Y_{n+1}$ with feature $\bm{x}_{n+1}$, let the PPD for $Y_{n+1}$ be the {\it Bayesian conformity score function}, scoring the plausibility of $z = (\bm{x}_{n+1}, y)$ in light of the augmented set $\D^y_{n+1} = \D_n \cup \{(\bm{x}_{n+1}, y)\}$, that is,
\begin{align} \label{eq: PPD_fullCP}
r^{\ppd}\big(z;\D^y_{n+1}\big)\;=\;p\big(y\mid\bm{x}_{n+1},\D^y_{n+1}\big).
\end{align}
The augmented set guarantees the exchangeability of the resulting scores and the validity of the full CP. Pragmatically, the procedure can be described as follows.

\begin{mybox}[frametitle={Procedure I. \emph{Bayesian Full CP}}]
\begin{description}
 \item[Input] (1) Sample $\D_n = \{Z_i = (\bm{X}_i, Y_i)\}_{i = 1}^n$ and (2) feature $\bm{x}_{n+1}$ of a new unit; (3) Working model $f(y|x, \theta)$ and (4) prior $\pi(\theta)$ on $\theta$; (5) Reasonable (possibly dense) grid $\mathcal{Y}^{\text{grid}}$ for $y$; (6) Coverage $1-\alpha,\,\alpha \in (0,1)$. 
  \item[Candidate evaluation] For each candidate $y \in \mathcal{Y}^{\text{grid}}$:
  \begin{enumerate}
      \item Let $z = (\bm{x}_{n+1}, y)$ and form the augmented dataset $\D^y_{n+1}=\D_n \cup z$.
      \item Compute the PPD of a future $Y_{n+1}$ on the augmented sample $\D^y_{n+1}$: $$ p\big(\cdot\mid\bm{x}_{n+1},\D^y_{n+1}\big).$$
      If not available in closed form, Monte Carlo (MC) methods can be adopted~\cite{robert2004monte}. 
      \item Get the conformity scores 
      using the PPD in (ii):
      \begin{align*}
      R_i& = p\big(y_i\mid\bm{x}_{n+1},\D^y_{n+1}\big), \quad i = 1,\dots, n,\\
      R^y_{n+1}& = p\big(y\mid\bm{x}_{n+1},\D^y_{n+1}\big).
      \end{align*}
      \item Evaluate the inclusion of candidate $y$ by comparing its score to the {\it conformal quantile} obtained from scores of the observed data ($\lfloor \cdot \rfloor$ is the floor function):
      \begin{align} \label{eq: conf_q}
      q_{n, 1-\alpha} \doteq R_{(\lfloor \alpha(n+1) \rfloor)}\ \text{among}\ \{R_i\}_{i =1}^n.
    \end{align}
      Include the candidate $y$ in the prediction set iff $R^y_{n+1} \geq q_{n, 1-\alpha}$. 
  \end{enumerate}
  \item[Output] Get the $100(1-\alpha)\%$ full CP interval $\mathcal{C}^{\text{full, PPD}}_{n, 1-\alpha}(\bm X_{n+1})$ for $Y_{n+1}$ as:
  \begin{align} \label{eq: CP_full_Bayes}
      \mathcal{C}^{\text{full, PPD}}_{n, 1-\alpha}(\bm X_{n+1}) = \{y\!: R^y_{n+1} \geq q_{n, 1-\alpha}\}.
  \end{align}
The set satisfies $\mathbb{P}({Y_{n+1}}\in \mathcal{C}^{\text{full, PPD}}_{n, 1-\alpha}(\bm X_{n+1}) ) \geq 1-\alpha$, regardless of the working model, the prior, and the sample size. 
\end{description}
\end{mybox}

\begin{remark}
    The reader may have noticed the difference between the conventional definition of the full CP in Eq.~\eqref{eq: CP_full} and its Bayesian version in Eq.~\eqref{eq: conf_q}-\eqref{eq: CP_full_Bayes}. This is because, conventionally, a ``non-conformity'' score function in the form of a residual is used; whereas, in the Bayesian use of the PPD, we are adopting a ``conformity'' measure: the higher the PPD at $y$, the more similar is $y$ to the sample data.  
\end{remark}

More than a decade ago, an equivalent version of the above Bayesian full CP was anticipated by Wasserman~\cite{wasserman2011}. Specifically, the suggestion was to repeatedly test (for a set of candidates $y \in \mathcal{Y}^{\text{grid}}$) the null hypothesis $H_0\!: Y_{n+1} = y$ using as the test statistic the PPD computed on the $y$-augmented sample. Then, a valid Bayesian prediction interval is obtained by inverting the test. In practice, the step {\bf Candidate evaluation}--(iv) is re-defined using a {\it conformal p-value} $p^y$ for $y$:
\begin{enumerate}
\item[(iv)] Evaluate the inclusion of candidate $y$ by computing its {\it conformal p-value}:
      \begin{align*} 
      p^y = \frac{1}{n+1} \sum_{i=1}^{n+1} \mathbb{I}(R_i \leq {R_{n+1}^{y}});
      \end{align*}
      Include the candidate $y$ in the prediction set if and only if $p^y\geq\alpha$. In fact, under $H_0$, all scores are exchangeable, implying that the rank of $R^y_{n+1}$ is uniformly distributed and therefore, a valid p-value.
\end{enumerate}
Using the test-inversion procedure, a final interval is then obtained as: 
$$\mathcal{C}^{\text{full}}_{n, 1-\alpha}(\bm X_{n+1}) = \{y\!: p^y\geq\alpha \}.$$

\begin{remark} \label{rem: deleted}
    We emphasise that the augmented dataset constructed as $\D^y_{n+1} = \D_n \cup \{(\bm{X}_{n+1}, y)\}$ guarantees exchangeability of the resulting scores $R_1,\dots, R_n, R^y_{n+1}$, though alternative (sometimes equivalent) formulations are also valid. The original version of full CP outlined in \cite{vovk2005algorithmic} was, in fact, based on a deleted (augmented) set $\D_{n+1,-i}^y \doteq \D^y_{n+1} \setminus \{(\bm{x}_i,y_i)\}$. The $i$-th omitted element is the one for which the conformity score is computed, i.e., $R_i = r\big(Z_i;\D^y_{n+1,-i}\big)$. The preference for one version over another is determined by the conformity score in use, as it may affect computational complexity, the efficiency of the resulting interval, and the possibility of undefined cases.
\end{remark}

\subsubsection{Efficiency and Bayes-risk optimality} \label{sec: Bayes_opt}

A prediction procedure is valid if it possesses frequentist coverage guarantees. In practice, we also want it to be precise or {\it efficient}; this can be quantified in terms of its expected volume: the smaller the better. Following this rationale, a Bayesian decision-theoretic justification for the use of a Bayesian conformity score function is given in \cite{hoff2023}. Specifically, this work shows that implementing full CP with the PPD leads to an efficient {\it Bayes-optimal} solution among all prediction methods that achieve the same (or higher) frequentist coverage. 

Let $\mathcal{P} = \{ P_\theta: \theta\in\Theta\}$ be a family of joint probability distributions for $Z = (\bm{X}, Y) \in \mathcal{X} \times \mathcal{Y}$. Denoted by $\lambda$ a volume measure on $\mathcal{Y}$, the risk function of a prediction procedure for $Y$ with $\bm{X} = \bm{x}$ is the expected size of its prediction set $\mathcal{C}(\bm{x})$ as a function of $\theta$; that is,
\begin{align*}
    \mathcal{R}_\theta(\mathcal{C}) = \mathbb{E}_{P_\theta}\big[ \lambda(\mathcal{C}(\bm{x})) \big]. 
\end{align*}
Clearly, a prediction region $\mathcal{C}$ is better than $\mathcal{C}'$ if $\mathcal{R}_\theta(\mathcal{C}) \leq \mathcal{R}_\theta(\mathcal{C}')$ and $\Prob_{P_\theta}(\mathcal{C}) \geq \Prob_{P_\theta}(\mathcal{C}')$ for all $\theta$, with inequality for some $\theta$. However, as in other statistical decision problems, typically, there is no uniformly best solution, motivating the search for optimality within a reduced class of procedures or in particular regions of the parameter space. 
For an example, see \cite{evans1980optimum}.

In many problems, there is indirect or prior information on distributions $P_\theta \in \mathcal{P}$ that are more likely than others. This suggests using a prior $\pi$ on $\Theta$, and evaluate the prediction region with a {\it Bayes-risk} function defined as
\begin{align} \label{eq: Bayes-risk}
    \mathcal{R}_\pi(\mathcal{C}) = \int_\Theta \mathcal{R}_\theta(\mathcal{C}) \, \pi(d\theta).
\end{align}
In this way, relevance is given to prediction regions that perform well for values of $\theta$ that are most plausible, at the expense of worse performance for less plausible values.

Extending Faulkenberry's \cite{faulkenberry1973method} result on the use of sufficient statistics for constructing valid prediction regions, Hoff \citep[][Theorem 3.2. and Theorem 3.4.]{hoff2023} shows that, for statistical models $\mathcal{P}$ possessing a boundedly complete regular
sufficient statistic, 
one can obtain a Bayesian prediction region $\mathcal{C}^\pi$ that jointly satisfies:
\begin{align*}
\Prob_{P_\theta}(\mathcal{C}^\pi) &\geq 1-\alpha \qquad \forall \theta \in \Theta, \tag{Frequentist validity}\\
\mathcal{R}_\pi(\mathcal{C}^\pi) &\leq \mathcal{R}_\pi(\mathcal{C}) \qquad \forall \mathcal{C}\ \ \text{with}\ \ \Prob_{P_\theta}(\mathcal{C}) \geq \Prob_{P_\theta}(\mathcal{C}^\pi) 
,\quad \forall \theta \in \Theta. 
\tag{Bayesian efficiency}
\end{align*}
That is, $\mathcal{C}^\pi$ has frequentist validity and is Bayes-risk optimal among prediction regions $\mathcal{C}$ with equal or greater coverage.

The construction is based on inverting the acceptance regions of a collection of conditional point-null hypothesis tests, where the conditioning element is a sufficient statistic. Notably, full CP is a special case of Faulkenberry's method, as applied to exchangeable models. In fact, for exchangeable data $Z_1,\dots, Z_{n+1}$, the unordered multiset of observations (the ``order statistics'') is a sufficient statistic. Faulkenberry's prediction method in this case is 
that, having observed $\D_n = \{Z_i = (\bm{X}_i, Y_i)\}_{i = 1}^n$, a candidate value $y \in \mathcal{Y}^{\text{grid}}$ is included in the prediction region for $Y_{n+1}$ if $y$ is in the acceptance region of a level-$\alpha$ test of $Y_{n+1}$ being uniformly distributed, 
as this is the conditional distribution of $Y_{n+1}$ given 
$\D^y_{n+1} = \D_n \cup \{(\bm{X}_{n+1}, y)\}$, for 
any member $P$ of $\mathcal P$. This is exactly the full CP procedure illustrated in Section 3~\ref{sec: Bayes_fullCP} using the inversion of the test advocated by Wasserman \cite{wasserman2011}. 

Importantly, for many models, the unordered values $\{Z_1,\dots, Z_{n+1}\}$ are not only a sufficient statistic, but a complete sufficient statistic. In particular, this occurs when $\mathcal{P}$ is the space of probability measures dominated by a common measure $\lambda$, where $\lambda$ is non-atomic (e.g., Lebesgue measure) or $\mathcal{Y}$ is countable. In these cases, Hoff \cite[][Theorem 4.1.]{hoff2023} shows that the Bayes-optimal choice of a conformity score is, not surprisingly, the Bayesian PPD, specifically, $p(Y_{n+1} \mid \bm{X}_{n+1}, \D_{n})$. A Bayes-optimal full CP procedure is obtained by using as conformity measure a deleted or {\it Leave-One-Out} PPD version, where $R_i = p\big(Y_i\mid\bm{X}_{i},\D^y_{n+1,-i}\big), i=1,\dots,n$, and $R^y_{n+1} = p\big(y\mid\bm{X}_{n+1},\D_{n}\big)$; see also Remark \ref{rem: deleted} and Section 3--(\ref{sec: computation}).

\subsubsection{Computational complexity} \label{sec: computation}

Bayesian full CP combines two major sources of computational complexity. The first arises from the Bayesian framework itself: beyond conjugate models, neither the posterior distribution nor the PPD is available in closed form. Asymptotically exact posterior samples can be obtained via sampling schemes such as Markov chain Monte Carlo (MCMC)~\cite{robert2004monte}, but this step can be computationally demanding, especially for large datasets or high-dimensional parameter spaces. The second, typically dominant, source of complexity stems from the full CP procedure. Here, one must re-fit the PPD for each candidate $y \in \mathcal{Y}^{\text{grid}}$ appended in the augmented data $\D^y_{n+1} \doteq \D_n \cup \{(\bm{X}_{n+1}, y)\}$. When using the deleted variant (see Remark~\ref{rem: deleted}), this burden increases further, as each fit must also exclude individual points. Consequently, the Bayesian computational cost scales linearly with the grid size $|\mathcal{Y}^{\text{grid}}|$ for the standard version, or $|\mathcal{Y}^{\text{grid}}| \times (n+1)$ for the deleted version, making a naive implementation of full Bayesian CP practically prohibitive.

Two solutions that address the computational burden in a full Bayesian CP framework are presented below. A first one, called {\it Add-One-In}, is implementable using the conventional PPD defined in Eq.~\eqref{eq: PPD_fullCP} based on the augmented set $\D^y_{n+1}$, while a second one, {\it Leave-One-Out}, can be adopted in the deleted-set version.
 
\paragraph{Add-One-In}
Consider the PPD in Eq.~\eqref{eq: PPD_fullCP} and let
    $m_r(\D_r) = \int_\Theta \prod_{i=1}^r f(y_i\mid \bm{x}_i,\theta)\, \pi(\theta) \, d\theta$
be the marginal distribution of the data $\D_r$. Given the augmented data $\D^y_{n+1}$, and using Bayes' rule and some algebra, we rewrite the PPD for a new sample $(\tilde y, \bm{x}_{n+1})$ as
\begin{align} \label{eq: ppd_trick}
p(\tilde y \mid \bm{x}_{n+1}, \D^y_{n+1})
&= \int_\Theta \frac{f(\tilde y\mid \bm{x}_{n+1},\theta) \, \prod_{i=1}^n f(y_i\mid \bm{x}_i,\theta)\, f(y\mid \bm{x}_{n+1},\theta)\, \pi(\theta) }{m_{n+1}(\D^y_{n+1})}\, {\dgray \frac{m_n(\D_n)}{m_n(\D_n)}} \, d\theta \nonumber\\
& = \frac{m_n(\D_n)}{m_{n+1}(\D^y_{n+1})}
  \int_\Theta f(\tilde y\mid \bm{x}_{n+1},\theta) \, f(y\mid \bm{x}_{n+1},\theta) \, \pi(\theta\mid \D_n) \, d\theta.
\end{align}

Similarly, one can notice that
\begin{align*}
m_{n+1}(\D^y_{n+1})
& = \int_\Theta f(y \mid \bm{x}_{n+1},\theta)\, \prod_{i=1}^n f(y_i\mid \bm{x}_i,\theta)\, \pi(\theta)\, d\theta = m_n(\D_n) \int_\Theta f(y \mid \bm{x}_{n+1},\theta)\, \pi(\theta\mid\D_n)\, d\theta,
\end{align*}
leading to
\begin{align} \label{eq: marg_ratio}
\frac{m_{n+1}(\D_{n+1})}{m_n(\D_n)}
= \int_\Theta f(y\mid \bm{x}_{n+1},\theta)\, \pi(\theta\mid\D_n)\, d\theta.
\end{align}

Both Eq.~\eqref{eq: ppd_trick} and~\eqref{eq: marg_ratio} use the same posterior distribution $\pi(\theta\mid\D_n)$; notably, this is uniquely based on the actual observed sample $\D_n$. Therefore, letting $\{\theta^{(g)}\}_{g=1}^G$ be a set of posterior draws from $\pi(\thetav\mid \D_n)$, we can easily approximate the PPD for the new sample $(\tilde y, \bm{x}_{n+1})$ as:
\begin{align} \label{eq: PPD_AOI_app}
   p(\tilde y \mid \bm{x}_{n+1}, \D^y_{n+1})\approx \frac{\sum_{g=1}^G f(\tilde y\mid \bm{x}_{n+1},\theta^{(g)})\,f(y\mid \bm{x}_{n+1},\theta^{(g)})}{\sum_{g=1}^G f(y\mid \bm{x}_{n+1},\theta^{(g)})} = \sum_{g=1}^G w^y_g\,f(\tilde y\mid \bm{x}_{n+1},\theta^{(g)}). 
\end{align}
This can be interpreted as a mixture of likelihoods $f(\tilde y\mid \bm{x}_{n+1},\thetav^{(g)})$ for $\tilde y$ with weights $w^y_g$ given by the specific contribution of the candidate $y$:
\begin{align*}
w^y_g = \frac{f(y\mid \bm{x}_{n+1},\theta^{(g)})}{\sum_{h=1}^G f(y\mid \bm{x}_{n+1},\theta^{(h)})}.
\end{align*}
Such weights approximate the ratio between the augmented posterior $\pi(\theta \mid \D^y_{n+1})$ and the original posterior $\pi(\theta \mid \D_n)$, enabling estimation of the PPD under the augmented dataset without re-fitting the posterior. We note that this approach is equivalent to the {\it Add-One-In Importance Sampling} strategy presented in \cite{fong2021}. 

\paragraph{Leave-One-Out} Consider the deleted version of the augmented sample set, $\D^y_{n+1,-i} \doteq \D^y_{n+1} \setminus \{(\bm{x}_i,y_i)\}$, and define the {\it Leave-One-Out} PPD for a new sample $(\tilde y, \bm{x}_{n+1})$ as
\begin{align} \label{eq: LOO_ppd}
p(\tilde y \mid \bm{x}_{n+1}, \D^y_{n+1, -i}) = \int_\Theta f(\tilde y\mid \bm{x}_{n+1},\theta) \, \pi(\theta\mid \D^y_{n+1, -i}) \, d\theta.
\end{align}
Using the same decomposition as in Eq.~\eqref{eq: ppd_trick}, we can re-express the PPD in Eq.~\eqref{eq: LOO_ppd} as:
\begin{align*}
p(\tilde y \mid \bm{x}_{n+1}, \D^y_{n+1, -i})
& = \frac{m_n(\D_n)}{m_{n+1}(\D^y_{n+1, -i})}
  \int_\Theta f(\tilde y\mid \bm{x}_{n+1},\theta) \, \frac{f(y\mid \bm{x}_{n+1},\theta)}{f(y_i\mid \bm{x}_i,\theta)} \, \pi(\theta\mid \D_n) \, d\theta,
\end{align*}
where
\begin{align} \label{eq: marg_ratioLOO}
\frac{m_{n+1}(\D_{n+1,-i})}{m_n(\D_n)}
= \int_\Theta \frac{f(y\mid \bm{x}_{n+1},\theta)}{f(y_i\mid \bm{x}_i,\theta)}\, \pi(\theta\mid\D_n)\, d\theta.
\end{align}

As in {\it Add-One-In}, this decomposition allows an efficient use of the posterior draws $\{\theta^{(g)}\}_{g=1}^G$ from $\pi(\theta\mid\D_n)$ to approximate the PPD as in Eq.~\eqref{eq: PPD_AOI_app}. In this case, the mixture is characterised by weights $w^{y, i}_g$ that depend on both the specific contribution of the candidate $y$ and that of the deleted sample $i$:
\begin{align} \label{eq: w}
w^{y, i}_g = \frac{f(y\mid \bm{x}_{n+1},\theta^{(g)})/f(y_i\mid \bm{x}_i,\theta^{(g)})}{\sum_{h=1}^G f(y\mid \bm{x}_{n+1},\theta^{(h)})/f(y_i\mid \bm{x}_i,\theta^{(g)})}.
\end{align}

Clearly, while the {\it Add-One-In} approach requires a single set of weights of size $|\mathcal{Y}^{\text{grid}}|$, the {\it Leave-One-Out} formulation requires computing a distinct set of size $n+1$ for each candidate $y \in \mathcal{Y}^{\text{grid}}$. While this comes at some computation and memory expenses, we emphasise that it still discounts the major burden of repeated MCMC operations. Importantly, {\it Leave-One-Out} can be advantageous in settings where {\it Add-One-In} degenerates, such as $k$-nearest neighbours (with $k=1$) or the noiseless Gaussian process, where the PPD is undefined (see Appendix B in \cite{fong2021}). Further, as discussed in Section 3--(\ref{sec: Bayes_opt}), this version has the property of being Bayes-risk optimal.

\subsubsection{Analytic solutions of full CP}

The computational complexity represents one of the main implementation challenges of a {\it full} CP framework, becoming even more prohibitive in a Bayesian setting. However, as illustrated in Section 3--(\ref{sec: computation}), the computational burden caused by the Bayesian component can be alleviated by the use of importance weights, which allow recycling the same posterior draws. This is possible for any (complex) model. Importantly, for certain models with closed-form posterior predictive distributions, full CP intervals can be obtained analytically. 

In this section, we discuss a very elegant approach to this problem, which involves the concept of \emph{equivalent conformity measures}~\cite[ECMs;][]{bersson2024}. An ECM is a conformity score function that preserves the same ordering in the scores as the original score function. Yet, replacing a computationally expensive score with an equivalent but simpler one can substantially reduce the computational burden without altering the resulting conformal region. For example, what if we could replace the {\it Leave-One-Out} PPD in Eq.~\eqref{eq: LOO_ppd}, which is Bayes optimal, by the conventional PPD?

\begin{definition}[{\bf Equivalent conformity measures};~\cite{bersson2024}, Definition 1]
    Two conformity measures $r$ and $s$ are called equivalent conformity measures if the resulting conformal prediction regions are equal. Following Lemma 6 in~\cite{bersson2024}, for given $z_i = (\bm{x}_i, y_i),\ i = 1,\dots,n$, and $z_{n+1} = (\bm{x}_{n+1}, y)$, if
    $$ \{y\!:r(z_i;\D)\leq r(z_{n+1};\D)\}=\{y\!: s(z_i;\D)\leq s(z_{n+1};\D)\}, \quad \text{for all}\ \ i=1,..., n, n+1,$$
then $r$ and $s$ are said to be equivalent conformity measures; we denote the ECM as $r \equiv s$.
\end{definition}

\paragraph{A continuous example: the Normal model \cite{bersson2024}} Consider an exchangeable sample $\D_n = \{ Y_i\}_{i=1}^n$, with unit outcomes $Y_i$ following a Normal distribution, with Normal and Gamma priors on its parameters (and known hyperparameters $\mu,\tau^2, a,b$); that is,
\begin{align}\label{workingmodel}
\begin{split}
 Y_i \sim{}& \text{Normal}(\theta,\variance),\quad i=1,\dots,n;\\
\theta\sim \text{Normal}(\mu,\tau^2\variance),&\quad \quad 
1/\variance\sim  \text{Gamma}(a/2,b/2).
\end{split}
\end{align}
For this Bayesian model, the PPD for a generic $\tilde y$ is a non-central Student-$t$ density:
\begin{align}\label{postpred}
p(\tilde{y}|\D_n) =
\frac{\Gamma \left(\frac{a_\sigma+1}{2}\right)}{\sqrt{a_\sigma\pi}\Gamma\left(\frac{a_\sigma}{2}\right)} 
\left(\frac{1}{\sqrt{\variance_t }}
\left( 1  +\frac{1}{a_\sigma}\frac{(\tilde{y}-\mu_\theta)^2}{\variance_t }
\right)^{-(a_\sigma+1)/2}\right),
\end{align}
where $\variance_t = b_\sigma(1+\tau^2_\theta)/a_\sigma$, with
\begin{align*}
a_\sigma = a+n,\quad\quad
b_\sigma = b + \sum_{j=1}^n y_j^2+\frac{\mu^2}{\tau^2}-\frac{\mu_\theta^2}{\tau^2_\theta},\quad\quad
\mu_\theta = \left(\frac{\mu}{\tau^2}+\sum_{j=1}^n y_j\right)\tau^2_\theta,\quad\quad
\tau^2_\theta = \left(\frac{1}{\tau^2}+n\right)^{-1}.
\end{align*}

A key result establishes that the PPD admits an ECM representation lying between the classical full CP, based on the augmented set, and its deleted version; that is,  $p(y_i \mid \D^y_{n+1}) \equiv p(y_i \mid \D^y_{n+1, -i})$ for all $i$'s~\cite[see][Theorem~1]{bersson2024}. The former can therefore be employed at a lower computational cost without trading off the Bayes optimality. Furthermore, as reported in Theorem~\ref{th: directCP}, its use allows for the derivation of a {\it closed-form} or {\it analytic} solution for the full CP interval.

\begin{theorem}[{\bf Analytic solution of Bayesian full CP under the Normal model}; \cite{bersson2024}, Theorem 2] \label{th: directCP}
    Under the working model in Eq.~\eqref{workingmodel} with PPD given in Eq.~\eqref{postpred}, if this is taken as the conformity measure, then the Bayes-optimal full CP interval $\mathcal{C}^{\text{full}}_{n,1-\alpha}$ can be computed analytically using the $k$-th and $(2n-k+1)$-th order statistic of $\boldsymbol v$; that is,
\begin{align*}
    \mathcal{C}^{\text{full}}_{n,1-\alpha} = \left( \boldsymbol v_{(k)}, \boldsymbol v_{(2n-k+1)}\right),
\end{align*}
with $k=\lfloor\alpha(n+1)\rfloor$ and $ \boldsymbol v = \begin{bmatrix}y_1&\cdots&y_n&g(y_1)&\cdots&g(y_n)\end{bmatrix}^T$, where
\begin{align*}
    g(y_i):= \frac{2\left(\mu/\tau^2+\sum_{j\in\{1:n\}} y_j \right)\left(1/\tau^2+n+1\right)^{-1} - y_i }{1-2\left(1/\tau^2+n+1\right)^{-1}}.
\end{align*}
Furthermore, the conformal prediction region is an interval that contains the posterior mean  $\hat{\theta}$.
\end{theorem}

\paragraph{A discrete example: the Binomial model} Consider now an exchangeable sample $\D_n = \{ Y_i\}_{i=1}^n$, with unit outcomes $Y_i$ following a Binomial model, with Beta prior for the success probability parameter $\theta$ (hyperparameters $a, b$ are assumed to be known); that is, 
\begin{align}\label{workingmodel2}
\begin{split}
 Y_i\sim{}& \mathrm{Binomial}(m_i, \theta),\quad i = 1, \dots, n,\\
\theta\sim {}& \mathrm{Beta}(a, b).
\end{split}
\end{align}
To preserve exchangeability, we shall consider a setting with $m_i = m$, for all $i = 1,\dots,n$ and $n+1$.

For this Bayesian model, the PPD for $Y_{n+1}$ is a Beta-Binomial distribution, where
\begin{align}\label{postpred2}
p(Y_{n+1} = \tilde{y}|\D_n) = \binom{m}{\tilde{y}}
\frac{B(\tilde{y} + a', m - \tilde{y} + b')}{B(a', b')},\quad \tilde{y}=0,\dots,m,
\end{align}
where $a' = \sum_{j=1}^n y_j + a$ and $b' = n m - \sum_{j=1}^n y_j + b$.
Under this model, it is straightforward to verify that the {\it Leave-One-Out} PPD does not have an ECM as in the Normal example. In fact, it is easy to show that $\exists i\ \text{s.t.}\ p(y_i | \D^y_{n+1}) \not\equiv p(y_i | \D^y_{n+1, -i})$, meaning that the two are not ECM.

\begin{theorem}[{\bf Equivalent Conformity Measures under the Binomial model}] \label{th: ECM_binom}
Consider the Binomial model in Eq.~\eqref{workingmodel2} and take as conformity measure the absolute residual from its PPD mean, which we term {\it Bayesian residual}; that is:
\begin{align} \label{eq: Bay_res}
    r^{\Bres}(y; \D^y_{n+1}) \doteq |y - \mathbb{E}_{Y \sim p(\cdot\mid \D^y_{n+1})}(Y)|.
\end{align}
Then, $r^{\Bres}(y_i; \D^y_{n+1}) \equiv r^{\Bres}(y_i; \D^y_{n+1,-i})$ for all $i$'s, meaning that the two are ECM.
\end{theorem}

\begin{theorem}[{\bf Analytic solution of Bayesian full CP under the Binomial model}] \label{th: directCP2}
    Under the working model in Eq.~\eqref{workingmodel2}, if the Bayesian residual in Eq.~\eqref{eq: Bay_res} is taken as the conformity measure, then the full CP interval $\mathcal{C}^{\text{full}}_{n,1-\alpha}$ can be computed analytically using the $k$-th and $(2n-k+1)$-th order statistic of $\boldsymbol v$; that is,
\begin{align*}
    \mathcal{C}^{\text{full}}_{n,1-\alpha} = \left( \boldsymbol v_{(k)}, \boldsymbol v_{(2n-k+1)}\right)
\end{align*}
with $k=\lfloor\alpha(n+1)\rfloor$ and $ \boldsymbol v = \begin{bmatrix}y_1&\cdots&y_n&g(y_1)&\cdots&g(y_n)\end{bmatrix}^T$, where
\begin{align*}
    g(y_i):= \frac{2 m \left(\sum_{j\in\{1:n\}} y_j + a\right)}{n m + a + b} - y_i.
\end{align*}
Furthermore, the conformal prediction region is an interval that contains the posterior mean $\hat{\theta}$.
\end{theorem}
Proofs of Theorem \ref{th: ECM_binom} and \ref{th: directCP2} are deferred to the Appendix \ref{appendix}.

\subsection{Bayesian Split CP}
Split CP is a widely used alternative to full CP due to its computational simplicity. Importantly, this approach eliminates the need for repeated re-fits at each candidate $y \in \mathcal{Y}^{\text{grid}}$, making it a valuable candidate in a Bayesian setting. Surprisingly, despite its simplicity, practical relevance, and popularity, we are not aware of any prior work on a \emph{Bayesian split CP} version, nor of any theoretical insights into its behaviour. We attempt such a formalisation here.

Similarly to the Bayesian full CP, we consider an exchangeable sample $\D_n$ of $n$ observations and aim to quantify the predictive uncertainty for a future outcome $Y_{n+1}$ with features $\bm{x}_{n+1}$. For an arbitrary working model and prior, a Bayesian predictive distribution can be derived and used to form a Bayesian conformity score function. In principle, the PPD itself could serve this purpose. However, as will become clear shortly, certain summaries such as the Bayesian residual in Eq.~\eqref{eq: Bay_res} provide a more convenient route for deriving an {\it explicit} CP interval.

\begin{mybox}[frametitle={Procedure II. \emph{Bayesian Split CP} -- using the \emph{Bayesian residual} score
}]

\begin{description}
\item[Input] (1) Sample $\D_n = \{Z_i = (\bm{X}_i, Y_i)\}_{i = 1}^n$ and (2) feature $\bm{x}_{n+1}$ of a new unit; (3) Working model $f(y|x, \theta)$ and (4) prior $\pi(\theta)$ on $\theta$; (5) Coverage $1-\alpha,\,\alpha \in (0,1)$.
\item[Split] Random partition of $\D_n$ into a training set $\D_{n_T}^{\texttt{Train}}$ and a calibration set $\D_{n_C}^{\texttt{Cal}}$ of sizes $n_T$ and $n_C=n-n_T$, respectively: 
\[
\D_n = \D_{n_T}^{\texttt{Train}} \cup \D_{n_C}^{\texttt{Cal}}, 
\quad \D_{n_T}^{\texttt{Train}} \cap \D_{n_C}^{\texttt{Cal}} = \emptyset.
\]

\item[Training] Use $\D_{n_T}^{\texttt{Train}}$ under the given working model and prior to derive and fit a Bayesian predictive model, e.g., the PPD $p(\cdot\mid \bm{X},\D_{n_T}^{\texttt{Train}})$. If not available in closed form, get a posterior sample $\{\theta^{(g)}\}_{g=1}^G$ from $\pi(\thetav\mid \D_n)$ and get an estimate $\hat{p}(\cdot\mid \bm{X},\D_{n_T}^{\texttt{Train}})$.
  
\item[Calibration] For each $Z_i \in \D_{n_C}^{\texttt{Cal}}$, compute conformity scores using, e.g., the \emph{Bayesian residual}:
\begin{align*} 
    R_i = r^{\Bres}(Z_i;\D_{n_T}^{\texttt{Train}}) = \big|Y_i - \hat \mu_{n_T}(\bm{X}_i) 
    \big|, \quad i\in \D_{n_C}^{\texttt{Cal}},
\end{align*}
  where $\hat \mu_{n_T}(\bm{x}) = \mathbb{E}_{Y\sim p(\cdot\mid \bm{x},\D_{n_T}^{\texttt{Train}})}(Y|\bm{x})$ denotes the posterior predictive mean for a unit with feature vector $\bm{x}$.\\
  Let $q_{n_C,1-\alpha}$ denote the $(1-\alpha)$ conformal quantile of the calibration scores:
  \begin{align} \label{eq: quant_split}
    q_{n_C,1-\alpha} = R_{(\lceil (n_C+1)(1-\alpha)\rceil)}.
  \end{align}

  \item[Output] Get the $100(1-\alpha)\%$ split CP interval $\mathcal{C}^{\text{split, B-res}}_{n,1-\alpha}(\bm{X}_{n+1})$ for $Y_{n+1}$ as:
  \begin{align} \label{eq: CP_split_Bayes}
    \mathcal{C}^{\text{split, B-res}}_{n,1-\alpha}(\bm{X}_{n+1}) 
    & = \{y\!: \big|y - \hat \mu_{n_T}(\bm{X}_{n+1})\big| \leq q_{n_C, 1-\alpha}\} \nonumber \\
    & = \Big[\hat\mu_{n_T}(\bm{X}_{n+1}) - q_{n_C,1-\alpha},\ 
           \hat \mu_{n_T}(\bm{X}_{n+1}) + q_{n_C,1-\alpha}\Big].
  \end{align}
  The set 
  has frequentist coverage, regardless of both the working model and prior. 
\end{description}
\end{mybox}

Although split CP is less efficient than full CP due to sample splitting, it is substantially more scalable in Bayesian settings: in fact, only a single fit based on $\D_{n_T}^{\texttt{Train}}$ is required, without the need to compute importance weights or ECM versions. However, the conformity score function may certainly impact the final CP efficiency; the more informative it is, the more efficient the interval is expected to be. Below, we outline some options, all grounded in a Bayesian framework. 

\begin{description}
\item[Posterior quantiles]
A popular conformity score in Split CP is based on the so-called conformalised quantile regression, which we turn now into a Bayesian version. For a given quantile $\tau \in (0,1)$, let $\widehat{Q}_{n,\tau}(\bm{x})$ denote the $\tau$-th posterior predictive quantile (or rather its estimate) computed on the $n$ samples. Then,
\begin{align} \label{eq: QRes}
    r^{Q\Bres}(z;\D) = \max\{\widehat{Q}_{n_T, \alpha/2}(\bm{x}) - y, y-\widehat{Q}_{n_T, 1-\alpha/2}(\bm{x}) \},
\end{align}
can be used as a score function, with the resulting CP interval expressed as
\begin{align*}
\mathcal{C}^{\text{split}, Q\Bres}_{n,1-\alpha}(\bm{X}_{n+1}) 
    = \Big[\widehat{Q}_{n_T, \alpha/2}(\bm{X}_{n+1}) - q_{n_C,1-\alpha}, \widehat{Q}_{n_T, 1-\alpha/2}(\bm{X}_{n+1}) + q_{n_C,1-\alpha}\Big],
\end{align*}
where $q_{n_C,1-\alpha}$ is the classical conformal quantile computed as in Eq.~\eqref{eq: quant_split}. This score naturally adapts to heteroscedasticity and skewness in the predictive distribution.

\item[PPD residual]  
We now propose a new conformity score that introduces a conceptual shift from residuals defined on the observation scale to residuals defined on the distribution scale. The idea is to measure deviations in the corresponding PPDs of the observed value and a reference value such as the MAP $\hat y^{\text{MAP}}$; that is,
\begin{align} \label{eq: DRes}
    r^{D\Bres}(z;\D) = \left |{p(y \mid \bm{x}, \D)} - {p(\hat y^{\text{MAP}} \mid \bm{x}, \D)} \right|.
\end{align}
This novel perspective reinterprets residuals as \emph{distributional discrepancies}, aligning with information-theoretic approaches using, e.g., the Kullback–Leibler divergence. As such, this score encodes richer probabilistic information about the expected excess surprisal of using $y$ instead of $\hat y^{\text{MAP}}$, covering both aleatoric and epistemic uncertainty.
\end{description}

It is worth noting that the conformity score in Eq.~\eqref{eq: DRes}, as well as the natural PPD score, does not yield an explicit Split CP interval of the form in Eq.~\eqref{eq: CP_split_Bayes}. Indeed, they all rely directly on the PPD, and identifying admissible values of $y$ requires solving the inequality
\begin{align*}
    \Big \{y: p(y \mid \bm{X}_{n+1}, \D_{n_T}^{\texttt{Train}}) \geq q_{n_C, 1-\alpha}\Big \},
  \end{align*} 
which may not be straightforward in practice. In such cases, a hybrid CP strategy can be adopted: first, Split CP is applied to obtain the conformal quantile $q_{n_C,1-\alpha}$; then, a grid search over $\mathcal{Y}^{\text{grid}}$ is performed to identify the set of candidate $y$ values satisfying the above inequality. This procedure corresponds to executing the {\bf Split, Training, Calibration} steps of Bayesian Split CP (Procedure II), followed by step (iv) of Bayesian full CP (Procedure I). 

\vspace*{-.2cm}
\begin{remark}
Split CP can be extended to cross-conformal~\citep{vovk2015cross} or jackknife+~\citep{barber2021predictive} by averaging or aggregating across multiple splits, thereby improving efficiency while preserving validity. In the Bayesian context, such extensions may still benefit from the simplicity and interpretability of the Bayesian residual in \eqref{eq: Bay_res}.
\end{remark}

\section{An Empirical Evaluation for the Beta-Binomial Model}
The proposed Bayesian CP procedures are now assessed in terms of their empirical performance, in terms of both coverage (validity) and size (efficiency) of the resulting intervals. As a data-generating process, we use the Binomial model in Eq.~\eqref{workingmodel2}. Specifically, we consider a sample size of $n = 20$ independent observations with $Y_i \sim \mathrm{Binomial}(m_i, \theta)$, for $i = 1,\dots, n$, with size $m_i = m = 20$ and success probability $\theta = 0.7$. The future outcome $Y_{n+1}$ is drawn from the same distribution. The choice of the Binomial (compared, e.g., to the Gaussian case) allows us to analyse the discrete and bounded-support setting, which may challenge non-parametric approaches.

We adopt a $\mathrm{Beta}(a=1/2,b=1/2)$ prior on $\theta$, corresponding both to the reference and the Jeffreys’ prior for the Binomial model. This choice is motivated by its desirable invariance property and its frequentist matching behaviour: Jeffreys’ prior is known to yield posterior credible intervals with approximately correct frequentist coverage even for moderate $n$, serving as a neutral benchmark.

\begin{figure}[H]
    \centering
    \includegraphics[width=1.02\linewidth]{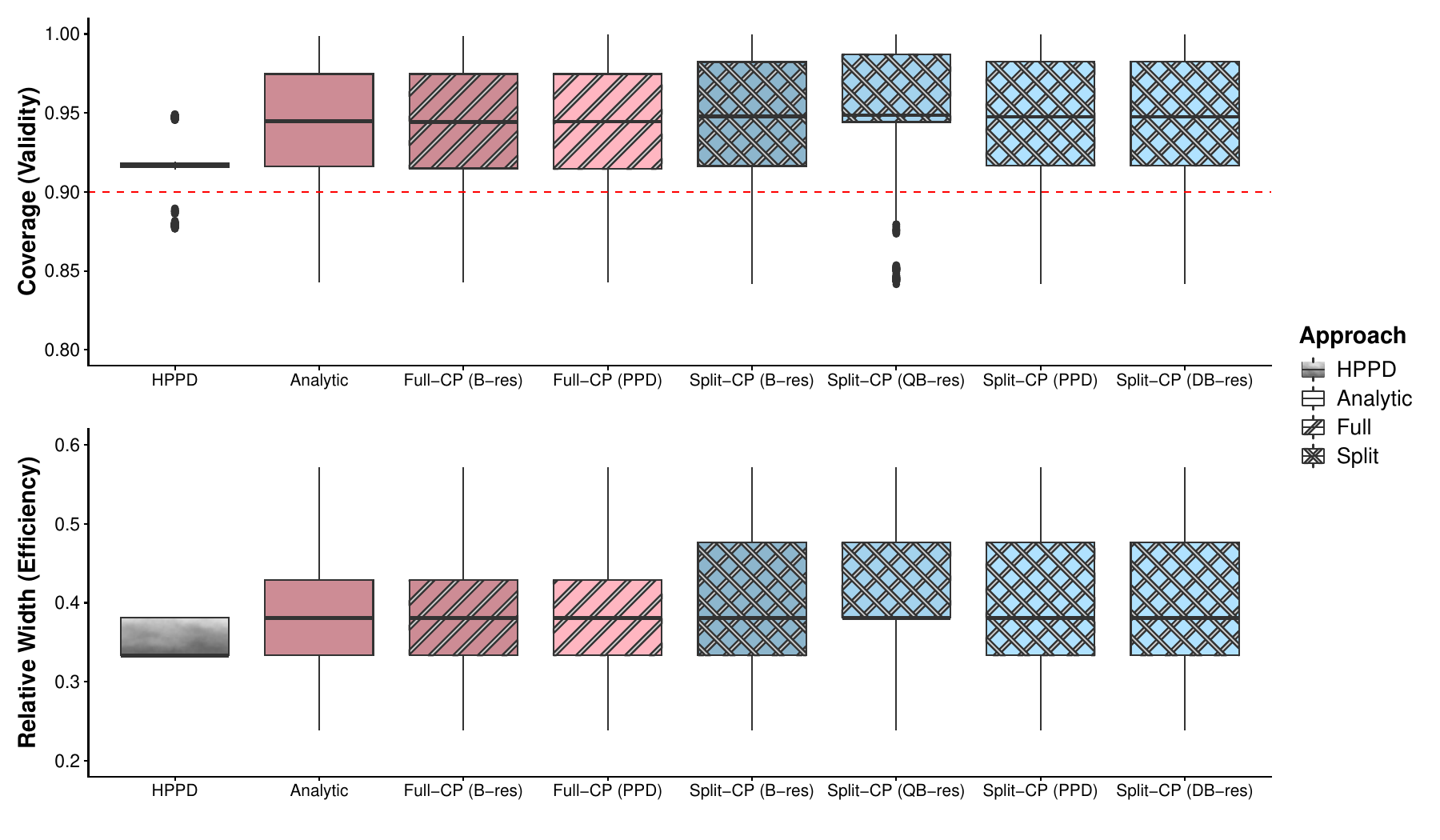}
    \caption{Empirical coverage and relative width (i.e., width relative to its full possible scale $m+1$) attained by the different Bayesian CP methods at nominal level 0.9 (red dashed horizontal line) under the Binomial model and the Jeffreys' prior. All results are based on 1000 MC replications.}
    \label{fig:covlen90}
\end{figure}

Prediction regions are constructed using both full and split Bayesian CP approaches, employing the range of Bayesian conformity measures discussed in Section~\ref{sec: CP_Bayes}: $r^{\ppd}$, $r^{\Bres}$, $r^{Q\Bres}$, $r^{D\Bres}$. The analytic solution derived in Theorem~\ref{th: directCP2} (corresponding to the full CP with $r^{\Bres}$), as well as the standard Bayesian HPPD interval, complement the evaluation. All results are based on $1000$ independent MC replications, with corresponding summaries provided in Figure~\ref{fig:covlen90}. These results confirm the theoretical properties of Bayesian CP in terms of frequentist coverage above the nominal level, here set as $1-\alpha = 0.9$. Among the CP variants, full CP and its analytic equivalent provide the best balance between validity and efficiency, with the latter providing a substantial computational advantage. Split CP versions exhibit greater variability and slightly reduced efficiency, as expected. Among the different Bayesian conformity measures, no significant differences emerge, except for $r^{Q\Bres}$, which tends to be more conservative. At first, Bayesian CP methods appear to offer no clear advantage over the HPPD benchmark. However, the Jeffreys' prior used here plays a pivotal role towards frequentist coverage. Moving from this noninformative prior toward stronger priors, the loss of frequentist coverage becomes evident.

\begin{figure}[H]
     \begin{subfigure}[b]{0.33\textwidth}
         \centering
         \includegraphics[width=\linewidth]{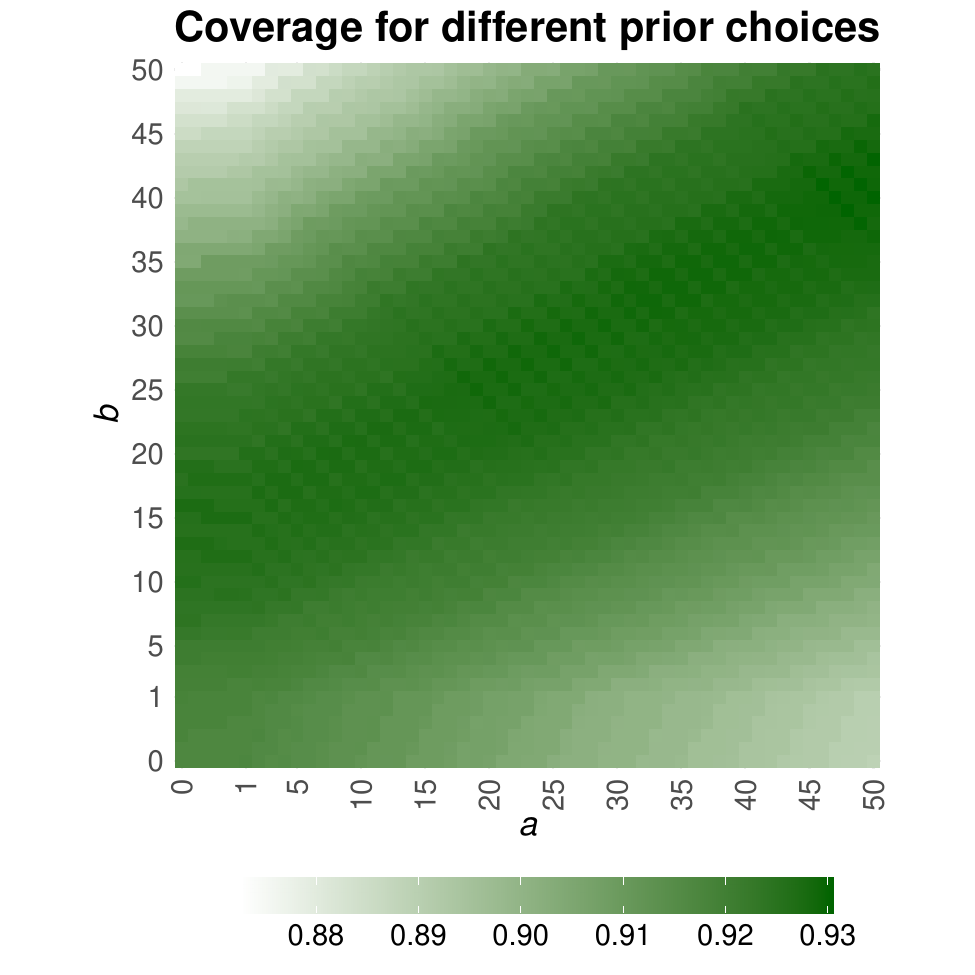}
         \caption{Bayesian HPPD}
         \label{fig: sq1}
     \end{subfigure}
     \begin{subfigure}[b]{0.33\textwidth}
         \centering
         \includegraphics[width=\linewidth]{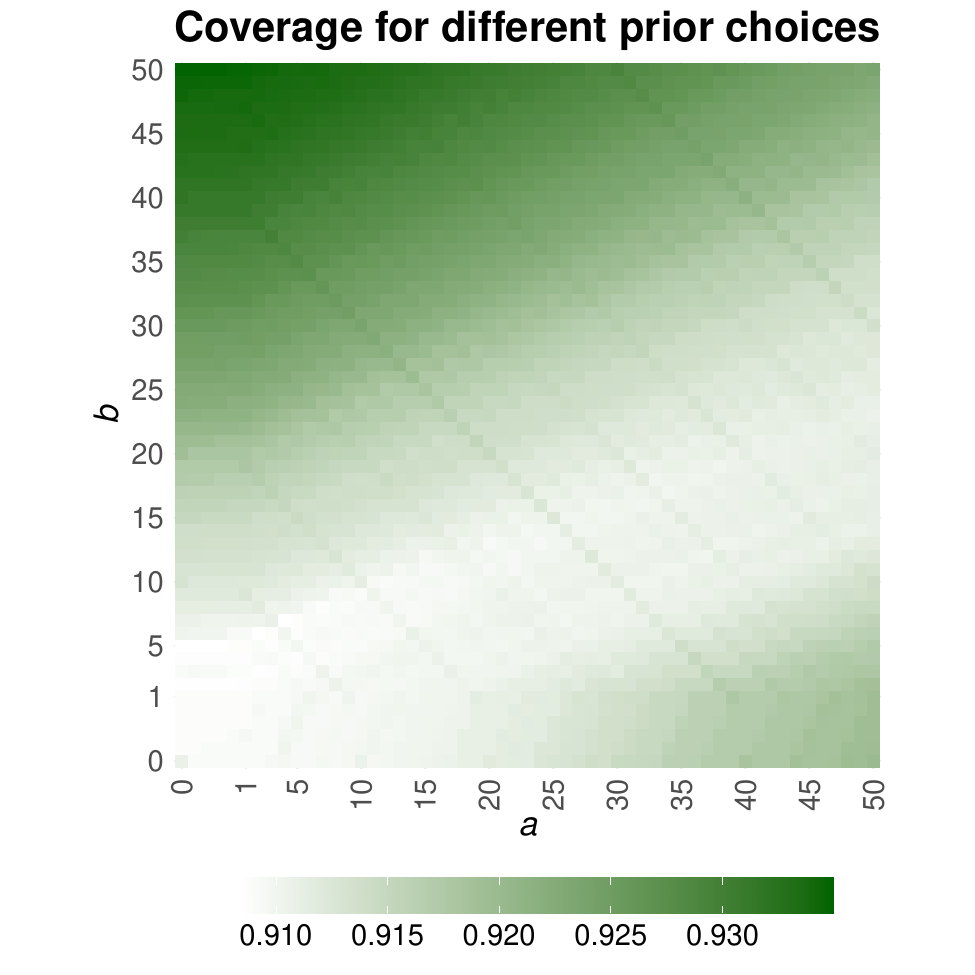}
         \caption{Bayesian CP (Analytic)}
         \label{fig: sq2}
     \end{subfigure}
     \begin{subfigure}[b]{0.33\textwidth}
         \centering
         \includegraphics[width=\linewidth]{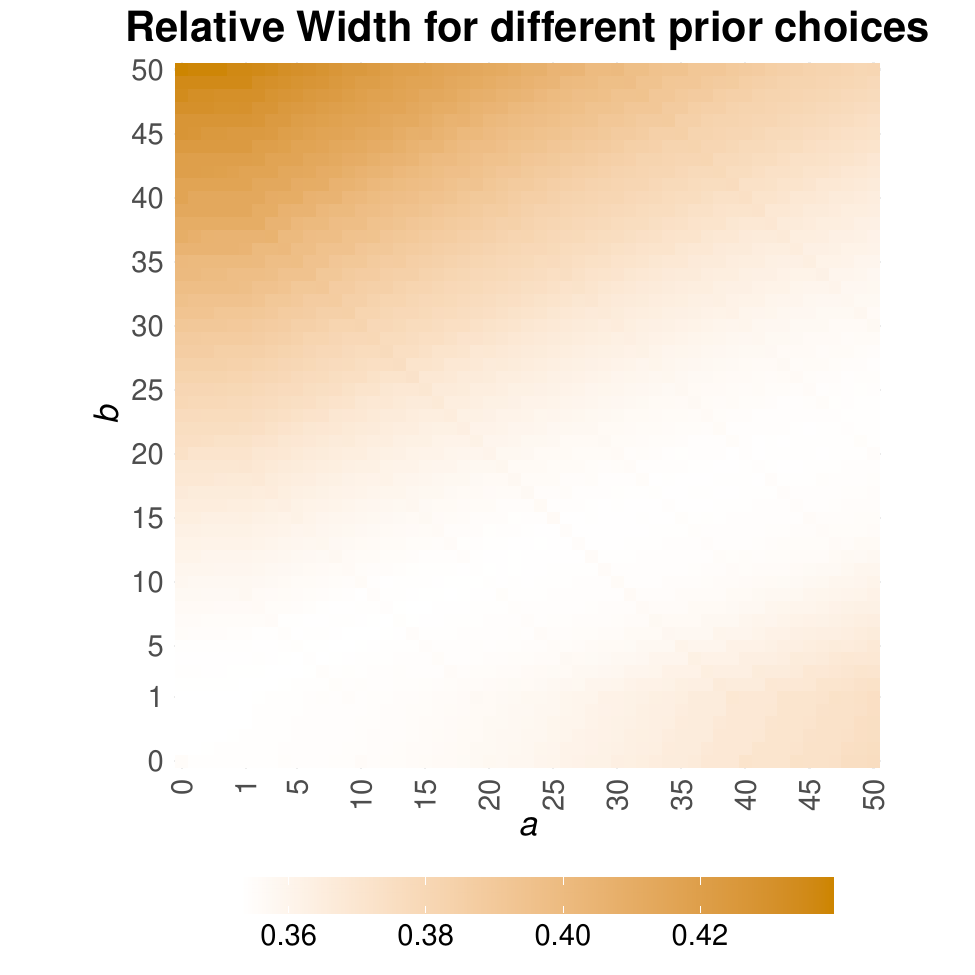}
         \caption{Bayesian CP (Analytic)}
         \label{fig: sq3}
     \end{subfigure}
        \caption{Role of prior hyperparameters $a,b$ in the classical Bayesian HPPD vs. Bayesian CP (Analytic full-CP with the Bayesian residual score $r^{\Bres}$). Nominal coverage is set to $0.9$.}
        \label{fig: prior}
\end{figure}

Figure~\ref{fig: prior} reports the coverage attained by both a classical HPPD interval and the analytical full-CP equivalent (with the Bayesian residual score $r^{\Bres}$), across different values of the prior parameters $a, b$. For strong priors (e.g. extreme hyperparameters $a,b$ for Beta), HPPD intervals fail to ensure nominal coverage (Figure~\ref{fig: sq1}), whereas Bayesian CP intervals remain above the nominal level, regardless of the prior (Figure~\ref{fig: sq2}). Yet, this comes at the cost of efficiency, as they prioritise validity through more conservative intervals. Efficiency (Figure~\ref{fig: sq3}) is highest when the prior aligns with the observed data (based on $\theta = 0.7$), with maximal values occurring for hyperparameter values yielding a prior mean near $0.7$. Thus, the Bayesian CP procedure adapts to prior-data agreement, producing narrower intervals when the prior agrees with the data while remaining robust under extreme (even conflicting) priors.

\vspace*{.1cm}
\section{Open Problems and Promising Directions}

\subsection{Conditional Coverage}
A well-known limitation of CP is that its finite-sample validity guarantees hold only \emph{marginally}, while \textit{conditional guarantees}, such as  
\begin{align} \label{eq: condit_cov}
\mathbb{P}\big(Y_{n+1} \in \mathcal{C}_{n,1-\alpha}(\bm{X}_{n+1}) \mid \bm{X}_{n+1}=\bm{x}\big) \geq 1-\alpha, \quad \forall \bm{x},
\end{align}
are in general unattainable without stronger assumptions. As a result, despite maintaining a target coverage rate on average across the covariate space $\mathcal{X}$, CP may undercover in certain regions of $\mathcal{X}$ while being overly conservative in others. Importantly, \emph{``the coverage can be quite poor for outlying groups, [...], 
which are likely the groups of highest concern''}~\cite{Hoff2022}.

In general, no method can provide conditional coverage, in a meaningful way, in a distribution-free setting~\citep{lei2014distribution}. Specifically, a prediction interval satisfying Eq.~\eqref{eq: condit_cov} is such that $\Prob\left (\lim_{\delta\to 0} \, \sup_{x\in B_\delta(x_0)} \lambda\left(\mathcal{C}_{n, 1-\alpha}(\bm{x})\right) = \infty \right )= 1$, for any non-atomic point $x_0$ of the distribution of $\bm{X}$. Analogous considerations can be done in a more general setting, such as linear regression~\citep{mccullagh2009conditional}.

Nonetheless, conformity measures that are more sensitive to local behaviour, including those in Eq.~\eqref{eq: QRes} and Eq.~\eqref{eq: DRes}, can achieve approximate conditional coverage. In this landscape, adopting a Bayesian framework based on PPD provides a promising way forward to bridge this gap. Indeed, the PPD allows for natural adaptation to heteroscedasticity, skewness, kurtosis, and multimodality, making it possible to construct intervals that resemble credible or highest-density regions~\cite{deliu2024alternative}. Compared to frequentist intervals, the latter also enjoy higher efficiency, in the sense that they tend to be shorter (see e.g., Section 7.1 in~\cite{ignatiadis2025empirical}).

The work of \citep{izbicki20a} first highlighted the potential of conformal methods driven by density measures, showing how the split CP can be combined with density estimators to yield more informative and efficient regions. Building on this, \citep{izbicki2022cd} established that such density-based conformal procedures not only guarantee marginal validity but also achieve stronger forms of validity, including local coverage and asymptotic conditional coverage. These contributions position the PPD as a principled mechanism for encoding richer distributional information and for moving closer to conditional coverage guarantees, while maintaining the finite-sample validity inherent to conformal prediction.

\subsection{Departures from Exchangeability}
The coverage guarantee of classical CP fundamentally relies on the assumption that data in $\D_n = \{Z_i=(\bm{X}_i, Y_i)\}_{i=1}^n$ are \emph{exchangeable}, meaning that, for any permutation $\pi$ of $\{1, \dots, n+1\}$,
$$(Z_1, \dots, Z_n, Z_{n+1}) \overset{d}{=} (Z_{\pi(1)}, \dots, Z_{\pi(n+1)}).$$

In practice, however, this assumption is often violated. Real-world data frequently exhibit grouped or hierarchical structures, temporal dependencies, or other forms of heterogeneity that break global exchangeability. For example, observations may be exchangeable within groups but not across groups, invalidating classical CP intervals if calibration and test points come from distinct exchangeability classes. Approaches to address this issue include {\it Mondrian}-style CP~\cite{vovk2005algorithmic}, which enforces local calibration within homogeneous subsets, or the adoption of a weaker form of exchangeability, referred to as {\it partial exchangeability}~\cite{diaconis1980finetti}. The latter has been adopted in~\cite{fong2021} under a hierarchical Bayesian framework, where different groups indexed by $j = 1,\dots, J$ are described by group-specific parameters $\theta_j$ partially pooled through a prior 
$\pi(\theta_j | \phi)$. Partial pooling allows information sharing across groups, enhancing the efficiency of each group-specific prediction interval.  

From a theoretical perspective, hierarchical Bayesian models are particularly appealing in this context, reflecting close connections with generalisations of de Finetti's representation theorem for partially exchangeable sequences~\cite{definetti38}. Therefore, the Bayesian perspective suggests that there is substantial potential for further integrating Bayesian ideas into CP for complex structured datasets, opening avenues for more flexible and theoretically grounded compromises.

\subsection{Small Area Estimation}

As outlined in Section~\ref{sec: intro}, a special setting of partial exchangeability where prediction is of extreme interest is found in {\it Small Area Estimation} (SAE)~\citep{rao2015small}. 
In this finite population framework, estimates and predictions are often needed at a granular level (small areas), yet the available sample is typically too limited or entirely absent to allow for precise {\it direct} estimation at refined spatial scales. Model-based approaches, including Bayesian hierarchical modelling, are therefore employed to borrow information strength across areas and improve estimation and prediction at the small-domain level. However, despite producing more precise estimates or (prediction) intervals, maintaining the desired coverage level in each area remains a challenge. We refer to~\cite{tzavidis2025, ranalli2025machine} for a modern account of these developments.

The introduction of CP methods into SAE, as explored in \cite{bersson2024}, provides a principled way to combine model-based inference with rigorous coverage guarantees. Importantly, CP methods in Bayesian SAE operationalise what Don Fraser~\cite{fraser2011} described as ``the obligation to study the frequentist properties'' of any statistical approach, ensuring that the reported coverage reflects actual reliability {\it in practice}, with implications for the status of statistics in science and society. Under general assumptions—such as approximate Gaussianity of the target variable and the availability of covariates only at the area level, Bersson and Hoff~\cite{bersson2024} show that CP methods can be applied directly, treating individual observations as approximately exchangeable within each domain. More challenging situations occur when the quantity of interest is a count and/or covariates are available at a unit level. Such settings pose open methodological problems that are currently under investigation and will be discussed elsewhere.

\paragraph{Acknowledgements} The authors wish to thank Larry Wasserman for helpful discussion. This work was supported by Sapienza grant n. $000041\_24$ Ateneo Medi 2023 – B83C2300691005 and Ateneo Piccoli 2024 – RP1241905F06F855.

\appendix

\section{Technical proofs} \label{appendix}

To prove {\bf Theorem~\ref{th: ECM_binom}}, we recall that under the working model in Eq.~\eqref{workingmodel2}, the PPD for $\tilde y$ with known $\tilde m$ (to preserve exchangeability, we must have $\tilde m = m$) is a Beta-Binomial distribution with parameters $a'$ and $b'$; that is,
\begin{align}\label{postpred22}
p(\tilde{y}|\D_n) = \binom{\tilde{m}}{\tilde{y}}
\frac{B(\tilde{y} + a', \tilde{m} - \tilde{y} + b')}{B(a', b')},
\end{align}
where $a' = \sum_{j=1}^n y_j + a$ and $b' = \sum_{j=1}^n m - \sum_{j=1}^n y_j + b$. Also, set $\sum_{j=1}^n y_j= n \bar{y}$.

Now, taking as the conformity score function the {\it Bayesian residual}, 
we have that:
\begin{align*}
    \mathbb{E} (Y \mid \D_n) = m\frac{a'}{a' + b'} = \frac{m\left(n \bar{y} + a\right)}{n m + a + b},
\end{align*}
and
$r^{\Bres}(y; \D_n) \doteq \left|y -\mathbb{E} (Y \mid \D_n) \right | $. 
Let $\D^y_{n+1} = \{ \D_n \cup y\}$ be the dataset augmented with the generic value $y$ and let $\D^y_{n+1, -i} = \D^y_{n+1} \setminus y_i$.  
We need to prove that the following holds:
\begin{align*}
    r^{\Bres}(y; \D^y_{n+1}) \equiv r^{\Bres}(y; \D_{n}),
\end{align*}
To prove this, we show that, $\text{for all}\ \ i=1,..., n$, and for all $y$,
$$ \{y\!:r^{\Bres}(y_i; \D^y_{n+1})\leq r^{\Bres}(y; \D^y_{n+1})\}=
\{y\!: r^{\Bres}(y_i; \D^y_{n+1, -i})\leq r^{\Bres}(y; \D_{n})\}.$$
This is equivalent to see that the following holds $\text{for all}\ \ i=1,..., n$ and all $y$:
\begin{align}
    r^{\Bres}(y_i; \D^y_{n+1}) / r^{\Bres}(y; \D^y_{n+1}) \leq 1 \quad \iff 
    \quad r^{\Bres}(y_i; \D^y_{n+1, -i}) / r^{\Bres}(y; \D_{n}) \leq 1.
\end{align}
However, in the Beta-Binomial framework
\begin{align*}
    r^{\Bres}(y_i; \D^y_{n+1, -i}) = \left \vert  y_i - m \frac{n \bar{y} + y - y_i+ a}{nm + a + b} \right \vert.
\end{align*}
Then, the following sequence of identical relations holds
\begin{align*}
   &\quad    r^{\Bres}(y_i; \D^y_{n+1, -i}) \leq  r^{\Bres}(y; \D_{n}) \\
   &\iff \left \vert  y_i - m \frac{n \bar{y} + y - y_i+ a}{nm + a + b} \right \vert
   \leq \left \vert  y - m \frac{n \bar{y}  + a}{nm + a + b} \right \vert \\
   &\iff \frac{\vert y_i (nm + a + b) - m (n \bar{y} -y_i + y +a \vert}
   {\vert y(nm + a+ b) - m (n \bar{y} + a)\vert } \leq 1 \\
   &\iff \frac{\vert y_i ((n+1) m + a + b) - m (n \bar{y} + y +a \vert}
   {\vert y((n+1) m + a+ b) - m (n \bar{y} + y + a) \vert } \leq 1,
\end{align*}
and this last relation actually implies that $r^{\Bres}(y_i; \D^y_{n+1}) \leq 
r^{\Bres}(y; \D^y_{n+1})$. $\blacksquare$

\noindent 
{\bf Proof of Theorem~\ref{th: directCP2}.} Let $\hat \theta = m \left ( n \bar{y} + y +a \right ) / \left ( nm + m +a + b\right )$.
The last inequality in the previous proof can then be restated as 
$$
\vert y_i - \hat{\theta} \vert \leq \vert y - \hat \theta \vert, \quad \quad \text{for all }i = 1, \dots ,n.$$ There are three possible cases, according to whether $y$ is smaller, equal or larger than $\hat \theta$, respectively. Simple algebra shows that 
$$
\begin{cases}
 y_i \in [y, 2\hat \theta - y] & y < \hat \theta \\
 y_i = y & y = \hat \theta \\
 y_i \in [2 \hat \theta, y] & y > \hat \theta \\ 
\end{cases}.
$$
Set $g(y) = 2 \hat \theta -y$. For each $i= 1, \dots, n,$, the range of values is 
$
(y \wedge g(y)) ; (y \vee g(y)).$
This implies that, for all $i = 1, \dots, n,$ the regions 
$$
S_i = \left \{ y : r^{\Bres}(y_i; \D^y_{n+1}) \leq r^{\Bres}(y; \D^y_{n+1})\right \}$$ are
intervals containing $\hat \theta$ and take the form
$$ (y_i \wedge g(y_i)) ; (y_i \vee g(y_i)).
$$
The full CP region will then be given by the $k$-th and the $(2n-k+1)$-th order statistics of the collection of bounds of the $S_i$'s. $\blacksquare$

\bibliographystyle{abbrv}
\nocite{*}
\bibliography{main}

\end{document}